# EVIDENCE OF CATALYTIC PRODUCTION OF HOT ATOMIC HYDROGEN IN RF GENERATED HYDROGEN/HELIUM PLASMAS

Jonathan Phillips,* Chun-Ku Chen, and Toshi Shiina
University of New Mexico, Department of Chemical and Nuclear Engineering,
Farris Engineering Center, Albuquerque, NM 87131.

## ABSTRACT

A study of the line shapes of hydrogen Balmer series lines in RF generated low pressure $H_2$/He plasmas produced results suggesting a catalytic process between helium and hydrogen species results in the generation of 'hot' (ca. 28 eV) atomic hydrogen. Even far from the electrodes 'hot' atomic hydrogen was predominant in $H_2$/He plasmas. Line shapes, relative line areas of cold and hot atomic hydrogen (hot/cold>2.5), were very similar for areas between the electrodes and far from the electrodes for these plasmas. In contrast, in $H_2$/Xe only 'warm' (<5 eV) hydrogen (warm/cold<1.0) was found between the electrodes, and only cold hydrogen away from the electrodes. Earlier postulates that preferential hydrogen line broadening in plasmas results from the acceleration of ionic hydrogen in the vicinity of electrodes, and the special charge exchange characteristics of $Ar/H_2^+$ are clearly belied by the present results that show atomic hydrogen line shape are similar for $H_2$/He plasmas throughout the relatively large cylindrical (14 cm ID x 36 cm length) cavity.

* To whom correspondence should be addressed. Phone: 505-665-2682; Fax: 505-665-5548; E-mail: jphillips@LanL.gov.

## INTRODUCTION

Recently, [1-17] unique and dramatic $H_\alpha$ line broadening has been observed in a variety of laboratories in specific mixed gas plasmas, particularly $Ar/H_2$ plasmas, operated at low pressure (milli-Torr) and generated, with microwave and glow discharge systems. It is universally agreed that the broadening must be Doppler broadening. Alternatives sources of line broadening such as Stark effect and instrument broadening are easily eliminated as they cannot explain the magnitude of the observed broadening, nor are these alternative explanations consistent with the lack of broadening observed for species, other than atomic hydrogen, found in the plasma. For example, the electron density required for Stark broadening of the magnitude observed is orders of magnitude higher than the density of atoms in the systems studied. In fact, in those cases in which charged particle density was measured, it was about four orders of magnitude less than the neutral density, quite consistent with expected charge species densities. This agreement on the source of the broadening creates a conundrum for the physics community. To wit: How can one neutral species in a mixed gas plasma have a temperature of approximately 300,000 K, while in the same plasma all other species, including electrons, have temperatures of less than 10,000 K?

Indeed, there is a debate regarding the mechanism of the selective heating of atomic hydrogen in these plasmas. One class of mechanisms can be called field acceleration (FA) models because, in all the various permutations of this class of models, it is necessary for the energy to be given to the hydrogen from the field. For example, it is postulated that ionic molecular hydrogen is accelerated in the cathode fall region, then is neutralized and picks up charge from Ar ions, and finally dissociates and emits [6-17].

To explain the symmetric broadening observed (both red and blue shift), these models require that an ideal reflector must be present (generally postulated to be an electrode) that has unity elastic reflectivity of hot neutral H atoms, and in addition fully randomizes the direction of the reflected species. As discussed later, these two requirements are in fact mutually exclusive.

The alternative model [7, 18, 19] is based on a new paradigm of quantum physics, Classical Quantum Mechanics (CQM). This paradigm, applied to the line broadening phenomenon [1-5], postulates that the energy of the H atoms is created by a non-field, 'chemical', process, that leaves at least one hydrogen atom called hydrino in a 'sub-ground' state. Moreover, the process that produces the 'sub ground state hydrogen', and concomitantly releases the energy to produce super heated 'standard' atomic hydrogen, requires a novel 'resonant transfer' interaction between hydrogen atoms and particular atomic catalysts, including $Ar^+$, $He^+$, O and/or previously created sub-ground state hydrogen.

The role of experimental physics in situations such as that described above is clear: Undertake programs that will produce distinctly different outcomes for competing hypotheses. Indeed, there are many published tests, spectroscopic [2, 18, 20-26], calorimetric [27, 28], NMR [29], etc. that indicate the CQM theory conforms to experiment rather better than standard quantum theory. Thus, there is a sound scientific basis for further tests, particularly relating to the ability of CQM and FA models to predict the outcomes of Balmer series line broadening in RF plasmas. Specifically, this study was designed to create dramatic differences in predicted outcomes between FA models and CQM hypotheses in an RF mixed $H_2$/He gas plasma system.

Several tests of the two theories were included. For example, line broadening was studied throughout the volume of $H_2$/He plasmas, and not just in the high field region between the electrodes. If any of the FA models are correct, then line broadening should not be observed in regions far from the high field region. In contrast, the effect of field on line broadening according to the CQM model should be secondary. That is, the intensity of the broadening might be lower away from the field, but the magnitude of broadening should be only marginally impacted. Also, if the CQM hypothesis is correct, line broadening should be observed as readily in a $H_2$/He mixture as it is in an $H_2$/Ar mixture. Also, in the regions away from reflector surfaces line broadening should be asymmetric. As the FA models indicated that Ar has an unusually large x-section for electron transfer, the effect in a helium plasma should be dramatically reduced. The experiments were successful, in that a simple summary is possible: *All outcomes were consistent with the predictions of CQM and inconsistent with all FA models.*

In addition to providing experimental data to distinguish between two theories, the data collected are of intrinsic significance. Indeed, $H_2$/He plasmas are among the most common structures of matter in the universe. If an interaction between these two gases in a plasma leads to the selective generation of extremely energetic hydrogen (not helium), even in the absence of a field, then this phenomenon needs to be thoroughly investigated.

## EXPERIMENTAL

Plasma Hardware- All plasmas were generated in a GEC cell [30] held at 0.4 Torr. This system consists of a large cylindrical (14 cm ID x 36 cm length) Pyrex chamber containing two parallel steel circular (8.25 cm diameter) plates, placed about 1 cm apart at the center (Figure 1a). RF power from a RF VII, Model RF 5 13.6 MHz power supply was sent to the plates through 8 mm diameter steel feeds, which entered the chamber through standard Ultratorr fittings, one on each end of the chamber. Gases, UHP grade (99.999%) He, $H_2$ and Xe, were metered into the chamber through Ultratorr fittings at one end, about 18 cm from the electrodes, using two mass flow controllers (MKS). The chamber was pumped using a Welch two stage rotary vane oil sealed vacuum pump (Model 8920) with a rated capacity of 218 l/min. This pump was attached to the chamber with a 1 cm ID Ultratorr fitting at the end opposite that at which gas entered. Pressure was measured with an MKS Baratron gauge place between the pump and the plasma chamber. All RF components of the system (chamber, power supply, and controller) were kept inside a copper mesh Faraday cage. All inner surfaces were painted black.

A precise mapping of the plasma was made possible by the use of a patterned light fiber holder placed directly above the plasma chamber, 35 cm X 15 cm in size, with drilled 0.2 cm in diameter holes one centimeter apart in a square pattern (Figure 1b). Positions listed below are relative to the (0,0) hole, which is the hole just above the center of the electrodes/center of the chamber. Thus for example, (-5, -4) is 5 cm 'behind' the electrodes and to the left, that is between the electrodes and the gas input and 4 cm to the left when the chamber is viewed from the gas exit port. In all cases, in which the light fiber was employed in the matrix, it was oriented vertically as shown in Figure 1b.

Spectrometer- The spectrometer system used in this study, described in detail elsewhere [31], is built around a 1.25 m visible light instrument from Jvon-Spex with a holographic ruled diffraction grating (1800 g $mm^{-1}$), with a nearly flat response between 300 and 900 nm. The slit was set at 10 μm in most cases. Light was collected using a light fiber bundle consisting of 19 fibers, each of 200 μm diameter. It is important to note that tests with a red laser with the system open clearly showed that light emanating from the region between the plates could not have reached the 'hooded' fiber optic probe when it was positioned at either end. Indeed, our tests showed that the readings are consistent with the listed 9° acceptance angle given with the probe. That angle indicates that a 1 cm diameter 'spot' would be encompassed by the acceptance cone even at the far side of the plasma cylinder from the probe.

In most cases, data was collected for the same time period, 10 one minute intervals. The repeat collection of data over the same spectral interval permits the system software to systematically eliminate 'noise spikes' by eliminating data 'points' that do not appear repeatedly. Thus, relative intensities in the figures provide very quantifiable comparison of the relative concentrations of excited hydrogen at any one collection point under different conditions (e.g. $H_2$/He ratios). They also provide a reasonable basis for a semi-quantitative comparison between excited H-populations at different collection points. Balmer series spectral lines were fit using two Gaussians, one for the fast hydrogen, and the second for the slow hydrogen using a procedure described elsewhere [31]. It is notable that the fittings achieved were excellent ($R^2$>0.97 in all cases).

**RESULTS**

The primary result of this work is the finding that in low pressure (< 0.4 Torr) RF generated $H_2$/He plasmas, in regions far from the strong field generated at the center of the electrodes, there is significant Doppler line broadening, indicating the dominance (Hot/Cold >2.5, Table I) of 'hot' hydrogen (> 20 eV) throughout the chamber. There is virtually no line broadening of the helium lines in these plasmas. These are the key findings consistent with the CQM predictions and contrary to the FA model predictions. Other findings that are apparently more consistent with the 'chemical process' CQM model are: (i) In $H_2$/Xe plasmas operated below 1 Torr, only cold hydrogen was found in regions away from the electrode, and only between the electrodes is there evidence of very small fraction of energetic, in this case 'warm' (<5 eV), hydrogen atoms (warm/cold <1.0). (ii) No line broadening of any species was found when the operating pressure exceeded 1.0 Torr. (iii) All of the Balmer lines are broadened to approximately the same energy. (iv) The impact of absorbed power and voltage on the magnitude of the line broadening was negligible. (v) The plasma is not symmetric. (vi) The relative intensity of the Balmer lines is not impacted by the position of the 'hot' electrode. In contrast, it is impacted by the position of the gas entry point.

In Figures 2a and 2b, a series of Balmer series lines collected at the gas exit end of the plasma system (Collection Point (15,3)) and from between the plates (Collection Point (0,0)) are shown for $H_2$/He plasmas with different $H_2$/He ratios for the 656.3 nm Balmer line ($H_\alpha$). The data in Table I confirms what is apparent upon inspection of this figure: the magnitude of the broadening is virtually the same at all collection locations. As shown in Figure 2c, upon substituting He for Xe, there is no hot hydrogen found anywhere in the system, although at high hydrogen concentration ($H_2$/Xe>4) a relatively small fraction of the hydrogen between the plates is 'warm'. At lower ratios only cold hydrogen is present consistent with previous explanations [1,4].

In Figure 3 - data taken about 2cm inward from the right 'edge' of the GEC cell: Collection Points (X, 5), it is shown that at high power (>150 W) the plasma is not symmetric. Indeed, the concentration of hot hydrogen is always higher near the gas entrance than at the pump end of the system. Reversing the 'ground' and 'hot' electrodes has no apparent impact. It is also clear that the magnitude of broadening is not a function of position along the axis, which is a clear indication that it is also not a function of electric field.

The tabulated data also makes it clear that the broadening is a function of many factors, but does not correlate with field strength. For example, the data in Table I make it clear that the broadening is a function of the ratio of helium and hydrogen. The data in this table also shows that the replacement of helium with xenon completely removes significant broadening, per Figure 2c. Table III data indicates that the strongest broadening is not always found in the highest field region, and also shows that the magnitude of the broadening is not a function of applied voltage. Indeed, the hottest hydrogen was recorded at 100 W, not at 150 W.

In Figure 4, spectra for another Balmer line ($H_\beta$), are shown for $H_2$/He plasmas from both ends of the system. This and other data (Table I) confirms an impression gained from visual inspection that the system is not symmetric. Also, from both figures it is clear that the Balmer lines for $H_2$/He plasmas actually consist of two parts, a broad base ('fast' hydrogen) and a sharp top ('slow' hydrogen).

This 'dual component' structure of the Balmer lines is not consistent with anything but Doppler broadening. Clearly it is not consistent with Stark broadening as that would affect all of the hydrogen, as well as the helium. Helium lines were consistently only about 0.01 nm in half width, or about equal to the instrument line broadening. Measurements of $H_2$/Xe plasmas produced only narrow Balmer series lines (<2.5 eV) away from the electrodes. Another interesting feature: The intensity increased with increases in the $H_2$/Xe ratio, clearly opposite to the trend observed in $H_2$/He plasmas. Only between the electrodes in $H_2$/Xe mixtures there is some warm hydrogen, up to 50% in one case, whereas in $H_2$/He plasmas the hot hydrogen is the dominant variety throughout the chamber.

Note (Table I and III) that the temperature of the 'hot' atomic hydrogen, determined from the $H_\alpha$ line width, is up to nearly 40 eV (i.e. Tmax > 400,000 K) at all locations. This is far greater than the 'temperature' ($T_{exe}$) of the atomic hydrogen (either the 'cold' or the 'hot') determined from a fit of the relative intensities using the Boltzmann method (approx. 0.5 eV or 5,000 K, determined at Position (15, -3), system operated at 100 W, with four Balmer lines). Moreover, in RF plasmas the excitation temperature is generally considered equal to the electron temperature; thus, the measured value of the atomic hydrogen is almost two orders of magnitude greater than that of the electrons [32-34].

The FA models also imply a relationship should exist between RF voltage and measured magnitude of line broadening. In order to probe this potential relationship the voltage/power relation was measured, as shown in Figure 6. Clearly the voltage between electrodes increases almost three times faster than the applied power in the region of interest, 40-150 W. Yet, no concomitant change in line broadening was discernable. In fact, as noted before, the greatest broadening was found at 100W input power, not at 150W.

## DISCUSSION

Many groups report (anomalous) line broadening of atomic hydrogen lines in mixed gas plasmas containing argon in high field regions [6-17]. A similar phenomenon has been observed in pure hydrogen plasmas, generated with DC discharge or RF systems including one group using a GEC cell [4, 7, 17]. These results repeatedly raise the same question: how can a plasma be created in which one particular neutral species is so much hotter than all other neutral species as well as so much hotter than the charged species? Below we argue that (i) all the existing standard physics models (FA models) fail to explain the data presented above, as well as other data available in the literature on this topic, and (ii) in contrast the CQM model is consistent with all observations.

Field Acceleration Models - As noted in the *Introduction*, all the FA models require that hot atomic hydrogen obtains its energy, directly or indirectly, from the field [12, 13]. For instance, in some models (gas phase production versions) the hot hydrogen is produced when a hydrogen ion, for example, $H_3^+$ [10], increased in concentration by interactions with Ar, is accelerated by the field toward an electrode, captures an electron via interaction with an Ar, electron dissociates to form n=3 state hydrogen, or forms n=3 state hydrogen via collision with a neutral Ar, and then emits. The possibility that the hot hydrogen forms from collisions with hot gas species (ions or atoms) is dismissed for two reasons. First, the cross sections for collisions with H ions are too small; second spectroscopy indicates there is little or no hot argon of any form present [7]. There are also models (surface production versions) in which the most energetic hydrogen atoms gain energy by 'bombardment' of hydrogen species on electrode surfaces with energized ions, generally $H_3^+$, $H_2^+$ or $H^+$ ions [6,7], and the subsequent ejection of hot hydrogen atoms [7].

In all cases, these gas production models only predict selective hydrogen broadening in high field regions, not throughout a large plasma chamber. Hot hydrogen ions created near the electrodes, the only region with sufficient field strength to accelerate ions to the measured energy levels, simply cannot migrate 15 cm without equilibrating with the system, even given a maximum computation of the mean free path of the hot hydrogen in a 300 mTorr system at 5000 K is about 0.5 cm. Moreover, these models require that the 'parent' species of the emitting H are hot ions. In this case, the excitation temperature, certainly after diffusing 15 cm, would be in equilibrium with the translation temperature, and would also be very high (i.e. >300,000 K); whereas, it was found to be only 5,000 K. Thus, any relevant model must account for a process that generates hot hydrogen atoms far from strong field regions. Indeed, the fields found 15 cm from the electrode are not nearly as strong as those found in the boundary layer near the electrodes. Field screening by the sheath reduces the fields dramatically within millimeters, and a highly conductive plasma bulk is essentially equipotential [7].

Another difficulty for the FA-gas models posed by the present results is the absence of argon in the present study. An 'adjusted' model would require that helium is also particularly good at generating hydrogen ions, and that there is rapid charge exchange between He, or some He ions, and $H_2^+$. Also, the evidence of relative populations clearly indicates no preferential population of n=3 states, contrary to the predictions of the FA gas models.

The 'distance' objection given to the FA-gas models also applies as an objection to the 'surface model' variations of the FA models. That is, these models are also inherently inconsistent with the observation of hot hydrogen many centimeters from an electrode surface. In this model class, hot H is presumably formed when high energy, field accelerated ions, generally $H_3^+$, $H_2^+$ or $H^+$ ions [6, 7] bombard an electrode surface rich in absorbed hydrogen, leading to the subsequent ejection of hot hydrogen atoms [14]. That is, energetic hydrogen created by electrode bombardment should only exist in a region of a few millimeters above the electrode, a region concurrent with the cathode fall region [7]. This result is to be expected as any 'hot' species will equilibrate, via collisions with other species, at a distance of no more than a few mean free paths of its creation point. Others suggest that surface generated hot hydrogen atoms are associated with a very broad peak, requiring the fitting of the overall line with three Gaussians [14]. This was not necessary in the present case (Figure 5).

Another major difficulty with any of the FA models is the fact that experiments universally show that viewed from any angle, the Balmer lines are symmetrically broadened. Moreover, there is no difference in the magnitude of the broadening as a function of the direction of the observation relative to the applied field. Thus, these models inherently require ideal reflectors that not only elastically scatter with perfect efficiency, but also reflectors that ideally scatter in all directions. Only such an ideal reflector could explain the remarkably even extent of blue and red shifts. (Yet, the two requirements are mutually exclusive, as discussed below.) This is generally recognized in the FA models, and often the electrodes are assigned these ideal properties.

Yet the two requirements of this proposed scattering process, perfect efficiency and 'randomization' of direction are mutually exclusive. Elastic collisions will not distribute the excess momentum, which is generated only in the direction of the applied field axis, in other directions, and only elastic collisions can create a symmetric broadening. Any inelastic processes would result in 'red shift', not blue, and hence 'randomization' processes should yield an asymmetric broadened peak. Moreover, the magnitude of broadening will decrease with each inelastic event. Thus, if inelastic events occur away from the electrode, as sometimes is suggested [17], they would produce steadily diminished broadening as a function of distance from the electrode. Clearly, this is not observed.

Standard Physics Alternatives- One alternative to the FA model that has been mentioned, but hardly developed, is the transfer of electron energy from electrons in the high end of the EEDF directly to the translational energy mode of atomic hydrogen [17]. This is clearly implausible as electron temperature is known to strongly couple to the atomic excitation temperature, and very weakly to translation temperature [35]. Thus, the measurement of excitation temperature in this study and in others of about 0.5 eV, is not consistent with electrons generating a translational motion temperature (of only one species!) almost two orders of magnitude greater than the excitation temperature.

Other conventional physics explanations for the observed line broadening also do not fit the values measured. For example, optical thickness cannot be a factor. Specifically, for optically thin plasmas (self adsorption not significant), the effective path length $\tau_\omega(L)$ is less than one:

$$\tau_\omega(L) = \sigma_\omega N_H L < 1 \tag{1}$$

where $\sigma_{\omega}$ is the absorption cross section, $N_H$ is the number density, and $L$ is the plasma path length traversed by the light. The absorption cross section for Balmer $\alpha$ emission is $\sigma = 1\,X\,10^{-16}\ cm^2$ [36]. An upper limit on the excited Hα density, assuming all of the hydrogen is fully dissociated, the temperature (as measured) is 5000 K, and, the excited states are populated according to the Boltzmann distribution (as measured), is $10^6/cm^3$. No more than 15 cm of plasma is traversed. These values indicate an effective path length of $10^{-8}$, clearly the plasma is optically thin for Hα, a result consistent with the observation that the Balmer α line is composed of both Doppler broadened and unbroadened components. A line broadened by self absorption would have only a single broad component. Stark broadening can also be eliminated. The required electron densities are orders of magnitude greater than the gas densities. Moreover, the lines are composed of hot and cold components. All H atoms, not just a fraction, as well as other species, would be impacted by high charge densities.

Even the Frank–Condon effect was considered. The 'Frank-Condon' effect [37-39] will create 'hot' neutrals with energies between 2 and 4.5eV via wall reactions of the type:

$$H_2^+ + e^- \rightarrow 2H \qquad (2)$$

Clearly, the energy of neutral species created in this fashion do not match the energies of the neutrals observed in this study.

One similarity between the results obtained here and earlier studies of $H_2$/Ar plasmas was the magnitude of the Hα broadening. In our case, for the strongest signal ($H_2$/He = $1x10^{-5}$) the measured broadening was 29 +/- 3eV, very close to the value recorded for $H_2$/Ar, which, for example, was more than 28 eV in recent studies [7, 17].

The results obtained in this study strongly imply that the energy required to create the Doppler broadening results from some form of catalytic process, involving a helium and hydrogen species, and not a process that requires energy absorption from the field, or field accelerated ion surface bombardment of electrodes, to produce hot H-atoms. The amount of energy released by the catalytic process must be remarkably high as more than 25 eV of energy is absorbed by the hot H atoms, far higher than the amount of energy generated (<1 eV/H atom) when water is formed from molecular hydrogen and oxygen.

Classical Quantum Mechanics (CQM) - The above requirements, based on observation made in this and other studies, are not consistent with the FA models, but they are consistent with the CQM model. In brief, the CQM model, based exclusively on Maxwell's equations and Newton's Laws, which includes the novel postulate that the electron as well as all other fundamental particles are three dimensional physical objects of precise shape, that follow classical physical laws, includes testable predictions of relevance to the current work. Specifically, it is postulated that the electron in the hydrogen atoms can undergo a transition from the 'traditional' ground state (principle quantum number, n=1) to a fractional quantum states called hydrinos with an energy far lower than that of the hydrogen that enters the reactor (e.g. $n = 1$ for which $E = -13.6$eV vs. $n = 1/2$, for which $E = -54.4$eV). These 'chemical' transitions of atomic hydrogen to stable states, can only take place due to the action of specific catalytic species, including $Ar^+$ and $He^+$. In this work, $He^+$ is clearly present in the plasma. In earlier line broadening studies, $Ar^+$ can be assumed to have been present. Also consistent with the theory is the finding that in those cases in which both Ar and He were excluded (i.e. Xe/H2 plasmas) no anomalous Balmer line broadening was observed.

The CQM model is also completely consistent with several other significant observations that pose difficulties for the FA model family. First, as the process that creates the excess energy is chemical, and not directly electric field driven, the hot H atoms produced should travel with equal likelihood in all directions. This will create a symmetric, Gaussian broadening, independent of direction of observation, as recorded herein. Second, also consistent with observation, the magnitude of the measured broadening should not correlate with applied field. Third, as the postulated process is chemical, the nature of the broadening should be the same everywhere. That is, there should be no clear trend, such as diminished magnitude of broadening with distance from the electrode. Again, observation is consistent with the predictions of the model.

In addition to the published spectroscopic studies cited above that are consistent with this model, it is noteworthy that calorimetric studies show that microwave generated $H_2$/He plasmas produce, for the same power input, approximately 50% more energy in a water bath than $H_2$/Xe plasmas generated with the same device under the same conditions of pressure, flow, etc. [27, 28].

It is important to note that in addition to the data presented here, there are several studies of plasma behavior designed to contrast FA model and CQM predictions that all clearly show the inadequacy of the FA models. For example, we also showed that in $H_2$/Ar plasmas produced in the same GEC cell, hot hydrogen is found throughout the chamber [40]. It has also been shown previously that selective heating of hydrogen atoms is found in a number of plasmas in which argon is not present [1, 3-5, 20-24]. Perhaps most compelling is a study showing that the phenomenon can be found in microwave-generated plasmas [24]. Field acceleration of ions, a requirement of the FA models, is not possible in microwave systems. Thus, the present data is consistent with that of several studies showing broadening in regions of plasmas that the FA models would predict no broadening. The mechanism of chemical transitions of atomic hydrogen will follow.

Mechanism of chemical transitions of atomic hydrogen by CQM - The average energy of the fast H, according to CQM, depends on the particular catalyst (resonant transfer or RT catalyst) as well as the conditions of the reaction. That is, as discussed below, the CQM theory indicates there are a number of processes that yield hot H atoms, with a range of energies. The average energy observed is a function of the relative rates of these processes, which occur in parallel and in series with each other. The rates of the individual processes are a function of catalyst identity and concentration, reaction cross sections, local (both time and space) concentrations of hydrinos, H atoms, etc. Below we provide an outline of the types of reactions which may occur, but it is not an exhaustive list. Also, no kinetic modeling was conducted, as that will require actual rate constants, as yet unavailable.

Below we develop a mechanism for selective heating of hydrogen, which is a application of earlier CQM mechanisms developed by R. Mills for $H_2$/Ar [40] and pure $H_2O$ plasmas [23]. According to the theory the helium plasma will initially (prior to build up of hydrino species) be dominated by an interaction between $He^+$ ions (the RT catalyst) and H atoms. The ionization of $He^+$ is approximately 54.4 eV and hence provides a net enthalpy of about 2, times $V_h$ (the potential energy of H being 27.2 eV or one hartree) and that qualifies it to be a catalytic species according to CQM. Specifically:

$$H(a_H) + He^+ \rightarrow He^{++} + H^*(a_H/3) + 54.4\ eV \tag{3}$$

where $H(a_H)$ is a 'ground state' hydrogen atom as conventionally understood, and $H^*(a_H/3)$ is a 'metastable' form of a $H(a_H/3)$ hydrino. $H(a_H/3)$ is a stable hydrino (shrunken atomic hydrogen) with a diameter 1/3 that of $H(a_H)$, and $H(a_H)$ has a radius of one Bohr radius.

This proposed reaction meets the postulated catalytic requirements of CQM. To wit: the process affecting the catalytic agent (i.e. ionization of the He ion) requires 2 × 27.2 eV which for this case is approximately half the energy of the transition (54.4 eV, per reaction (3)) to the lower (smaller) state of the hydrogen electron (108.8 eV). Hence, the postulated catalytic processes invariably create metastable hydrino species with a large 'excess' of energy. In particular, the metastable form of hydrino created in reaction (3) carries 54.4 eV of 'excess' energy. This is quickly lost via one of the following reactions:

$$H^*(a_H/3) \rightarrow H(a_H/3) + h\nu \tag{4}$$

or

$$H^*(a_H/3) + H(a_H) \rightarrow \text{Fast } H(a_H/3) + \text{Fast } H(a_H). \tag{5}$$

The former will create a 54.4 eV photon and the latter a fast hydrogen with a translational energy of 27.7 eV. Why is the fast photon emitted with twice the energy transferred to the atomic hydrogen? The reaction must conserve linear momentum. As written (eq. 5) both species enter the process with virtually no momentum, and thus must leave the reaction point with equal and opposite momentum. As both species have the same mass, they must also leave with equal energy.

It is interesting to note that without exception the data from the GEC shows that ALL of the energy absorbed by the $H(a_H)$ is absorbed as translational as predicted. That is, the proposed process as written is consistent with the observation that the excitation temperature of the hot H atoms observed is orders of magnitude lower than the translational energy.

Other processes that produce fast H atoms involve the reaction of one hydrino with another:

$$H(a_H/3) + H(a_H/3) \rightarrow H(a_H/2) + H(a_H/4) + 27.2 \text{ eV} \quad (6)$$

Where the 27.2 eV can either be given off as kinetic energy (e.g. reaction (5)) or as a photon (reaction (4)). The above reaction is representative of a set of processes termed 'disproportionation' [21, 27, 41]. These reactions are anticipated in the CQM as the hydrinos themselves have appropriate 'energy holes' to act as catalytic agents. Clearly starting with any set of hydrinos, produced catalytically per reactions such as Reaction (3), it is anticipated that lower energy state hydrinos, down to the lowest state (H(1/137)), can be produced via disproportionation processes. Each step 'down' to a smaller (literally) hydrogen species releases more energy, thus, the amount of energy available for creating high kinetic energy hydrogen atoms gets very large. Clearly, the relative concentrations of different hydrino species will change with time, and this can change the average energy observed. The hydrino concentration can also be expected to be position dependent. For example, detailed modeling might be used to test the postulate that hydrino concentration is highest at the pump end of the GEC cell.

Another set of processes involving two reacting hydrinos, can create hydrogen atoms. For example:

$$H(a_H/3) + H(a_H/2) \rightarrow H(a_H) + H(a_H/4) + 54.4\ eV \qquad (7)$$

It is possible that the metastable hydrino decomposes quickly enough that the $H(a_H)$ produced in reaction (7) absorbs 27.2 eV of kinetic energy. Moreover, if the plasma contains a high density of 'hot' hydrinos, likely given the metastable decomposition processes posited (e.g. reaction (6)), hydrino/hydrino reactions (e.g. reaction (7)) will create metastables of very high energies that upon collisional decay can produce atomic hydrogen with translational energies far higher than 27 eV. Such reactions can readily produce H atoms with both low (ca. 10 eV) and high energies.

On the basis of observation we offer a three-part answer to the question of why there is an obvious preference for the production of hot H atoms and not hot helium. First, the cross section for metastable decay to produce hot species via 'collision' with $H(a_H)$ is much higher than that with any other species. The experimental evidence clearly indicates the reaction, with H atoms is dominant. Possibly the cross section for metastable decay via conversion of energy to translational energy is only significant for this type of H*–H collisions. Collisions of the metastable with all other species, we suggest, would be unfavorable, and the drop in energy to the true stable state in the absence of atomic hydrogen will be via photon emission. Second, energy-transferring collisions between metastables and heavier species, with a concomitant drop in energy to the stable state, even if they do occur, will not produce a 'visible' effect. In a collision with a 27 eV hydrino, a He atom would absorb at most 5.5 eV. Moreover, the increase in velocity of species, and hence the relative change in line width goes as the square root of the mass ratio. Thus, relative to hydrogen, the absolute increase in line width of helium (assuming the same emission frequency) is only one-half. Thus, the net line broadening for He, even in the very unlikely event that the fraction of helium ‘heated’ by collision with metastable species was equal to that of H atoms, would be equivalent to the line broadening observed for 2.5 eV hydrogen. In all likelihood (see Reason 1, above) a tiny fraction of the He is heated in this way and hence the signal from this heating is lost in the averaging process. Third, inelastic collisions would not produce visible results. ‘Superheating’ of internal modes, such as electron excitation, would only be found in the EUV spectra, and hence could not be found with our system.

Direct (not photon mediated) transfer of energy from metastable species to atomic species is not a postulate unique to CQM theory, hence using the postulate herein is consistent with the general understanding of plasma physics at this time for another system, excimer H energy transfer. Selective and direct transfer of energy from metastable-noble-gase dimers (helium or neon) to hydrogen atoms to form excited gases [42-44] is suggested to explain several observations in these mixed gas systems. No line broadening is observed, and the ‘resonant energy transfer’ of energy is needed to explain clear evidence of selective excitation of one electronic transition in the hydrogen. For example, near atmospheric pressure admixture of Ne with very low hydrogen concentrations, excited with ionizing particle beams, results in very intense Lyman $\alpha$ radiation, with no concomitant Balmer lines, etc. [42]. This suggests a 10.2 eV resonant energy transfer from excited $Ne_2^*$ to H atoms, with the dominant cross-section for this particular process. As discussed in these reports, the excitations observed are at the far end of what is energetically possible. The observations in this report of hydrogen kinetic energies in excess of 30 eV are far outside the energy range achievable via the ‘metastable noble gas to hydrogen’ mechanism.

**SUMMARY**

The experiments conducted in this work were designed to distinguish two models of the widely reported selective superheating of H atoms in mixed gas, RF generated, plasmas. The experimental design was found adequate for that purpose. Specifically, the findings that (i) super heating of the atomic hydrogen is found in a H2/He plasma, (ii) the magnitude of the heating is virtually unchanged throughout a large GEC cell, (iii) the heating is not correlated to the field strength between the electrodes, or local field strength, (iv) only the translational energy mode is superheated (i.e. not the excitation energy), (v) the electrons are orders of magnitude cooler than the neutral H atoms, and (vi) the line broadening is symmetrical, independent of direction relative to the electric field direction are all totally inconsistent with the FA models and totally consistent with the predictions of the CQM model. In addition to the theoretical implications, the experimental results alone raise questions regarding processes taking place in gaseous $H_2$/He plasmas, one of the most common forms of matter in the universe (e.g. solar corona).

ACKNOWLEDGMENTS Dr. Randell Mills for many helpful discussions, particularly clarification of the CQM mechanism of selective H-atom heating in $H_2$/He plasmas.

Table I. Line Broadening Summary Balmer Series- He/$H_2$ plasma generated in the GEC cell operated at 0.4 Torr, constant power (100 W), and variable gas ratio.

| | ***(He + $H_2$)*** | ***Plasma*** | ***100W*** | |
|---|---|---|---|---|
| **Position / He Line** | **Flow Ratio ($H_2$/He)** | **Temperature Cold He (eV)** | | |
| 15,-3 / 667.76 (nm) | $1x10^{-5}$ | 0.155 | | |
| **Position / Balmer Lines** | **Flow Ratio ($H_2$/He)** | **Temperature Cold H (eV)** | **Temperature Hot H (eV)** | **Area Ratio (Hot/Cold)** |
| -15,-3 / $H_\alpha$ | $1x10^{-5}$ | 0.134 | 29.6 | **4.7** |
| -15.-3 / $H_\beta$ | $1x10^{-5}$ | 0.206 | 35.1 | **4.8** |
| 0,0 / $H_\alpha$ | $1x10^{-5}$ | 0.126 | 32.2 | **3.9** |
| 15,-3 / $H_\alpha$ | $1x10^{-5}$ | 0.125 | 26.8 | **4.0** |
| 15,-3 / $H_\beta$ | $1x10^{-5}$ | 0.166 | 29.1 | **4.9** |
| 15,-3 / $H_\chi$ | $1x10^{-5}$ | 0.243 | 28.7 | **3.9** |
| 15,-3 / $H_\delta$ | $1x10^{-5}$ | 0.239 | 21.1 | **2.5** |
| 0,0 / $H_\alpha$ | $1x10^{-5}$ | 0.126 | 32.2 | **3.9** |
| 0,0 / $H_\alpha$ | 0.4 | 0.141 | 25.6 | **2.4** |
| 0,0 / $H_\alpha$ | 0.8 | 0.137 | 23.7 | **2.1** |
| 0,0 / $H_\alpha$ | 1.2 | 0.138 | 23.3 | **1.6** |
| -15,-3 / $H_\alpha$ | $1x10^{-5}$ | 0.125 | 26.8 | **4.0** |
| -15,-3 / $H_\alpha$ | 0.4 | 0.153 | 33.8 | **3.1** |
| -15,-3 / $H_\alpha$ | 0.8 | 0.166 | 32.3 | **3.0** |
| -15,-3 / $H_\alpha$ | 1.2 | 0.173 | 29.5 | **2.8** |
| -15,-3 / $H_\alpha$ | 1.6 | 0.171 | 27.0 | **2.5** |
| -15,-3 / $H_\alpha$ | 2.0 | 0.169 | 24.8 | **2.1** |
| -15,-3 / $H_\alpha$ | 3.2 | 0.164 | 23.3 | **1.9** |
| -15,-3 / $H_\alpha$ | 4.8 | 0.158 | 19.5 | **1.7** |
| -15,-3 / $H_\alpha$ | 8.0 | 0.156 | 17.8 | **1.4** |

*Note: The temperatures were calculated using the following equation: $\Delta\lambda_{FWHM} \approx 7.1626 \times 10^{-7} \lambda(nm)\sqrt{T(^oK)}$ [5, 24]

Table II. Hα Line Broadening for Xe/$H_2$ Plasma- Plasma run at constant pressure (0.4 Torr), constant power (100 W), variable gas ratio.

| | ***(Xe + $H_2$)*** | ***Plasma*** | ***100W*** | Ratio |
|---|---|---|---|---|
| | **($H_2$ / Xe)** | **Cold H** | **Warm H** | Warm/Cold |
| 0,0 / $H_\alpha$ | 8.0 | 0.156 | 4.1 | **1.1** |
| -15,-3 / $H_\alpha$ | 3.2 | 0.121 | 2.5 | **0.3** |
| -15,-3 / $H_\alpha$ | 4.8 | 0.120 | 2.4 | **0.3** |
| -15,-3 / $H_\alpha$ | 8.0 | 0.120 | 2.4 | **0.3** |

TABLE III. Mapping $H_\alpha$ Line Broadening- Plasmas run at constant gas ratio (He/H2, 9/1) and pressure (0.5 Torr), and Hα broadening measured as a function of power and position.

| | | | 40W | | | 100W | | | 150W | | |
|---|---|---|---|---|---|---|---|---|---|---|---|
| Position | | | Temperature | | Area | Temperature | | Area | Temperature | | Area |
| ID | X [cm] | Y [cm] | Cold H [eV] | Hot H [eV] | Ratio (Hot/Cold) | Cold H [eV] | Hot H [eV] | Ratio (Hot/Cold) | Cold H [eV] | Hot H [eV] | Ratio (Hot/Cold) |
| 1 | 0 | 0 | 0.154 | 27.2 | 1.7 | 0.165 | 37.2 | 3.4 | 0.157 | 39.6 | 3.7 |
| 2 | -16 | 5 | 0.164 | 19.8 | 1.8 | 0.167 | 25.2 | 3.1 | 0.153 | 15.3 | 1.1 |
| 3 | -8 | 5 | 0.157 | 18.9 | 1.9 | 0.162 | 25.7 | 4.9 | 0.145 | 23.9 | 4.2 |
| 4 | 0 | 5 | 0.144 | 20.4 | 0.8 | 0.146 | 29.1 | 2.0 | | | |
| 5 | 8 | 5 | 0.155 | 14.6 | 1.5 | 0.153 | 24.7 | 5.6 | 0.148 | 21.9 | 4.0 |
| 6 | 16 | 5 | 0.155 | 25.0 | 1.1 | 0.162 | 25.6 | 2.6 | 0.161 | 22.2 | 2.3 |
| 7 | -8 | 3 | 0.158 | 19.6 | 1.8 | 0.147 | 26.0 | 4.7 | | | |
| 8 | 0 | 3 | 0.148 | 23.1 | 1.3 | 0.147 | 33.5 | 3.6 | | | |
| 9 | 8 | 3 | 0.156 | 15.4 | 1.5 | 0.160 | 24.1 | 4.9 | | | |
| 10 | -16 | 1 | 0.160 | 30.1 | 1.5 | 0.165 | 39.9 | 3.5 | | | |
| 11 | -8 | 1 | 0.159 | 27.8 | 1.6 | 0.163 | 34.1 | 4.3 | | | |
| 12 | 8 | 1 | 0.158 | 19.2 | 1.5 | 0.162 | 28.7 | 4.8 | | | |
| 13 | 16 | 1 | 0.155 | 26.7 | 1.5 | 0.164 | 35.0 | 3.5 | | | |
| 14 | -8 | -5 | 0.158 | 19.0 | 1.9 | 0.160 | 25.3 | 4.4 | | | |
| 15 | 0 | -5 | 0.144 | 20.0 | 0.9 | 0.155 | 31.6 | 3.3 | | | |
| 16 | 8 | -5 | 0.158 | 15.2 | 1.4 | 0.160 | 23.6 | 4.7 | | | |

## Figure Captions

Figure 1. GEC System. (a) Schematic of GEC cell. This entire cell is enclosed in a copper Faraday cage which is painted black. (b) A picture of the 'grid' employed to hold the fiber optic probe. In the picture the probe is placed at Position (0,0).

Figure 2. Balmer series $H_\alpha$ line broadening. This figure shows how the lines broadening changes as a function of position, $H_2$/He ratio and gas mixture. (a) Broadening measured at Position (-15,3) for $H_2$/He mixture; (b) Broadening measured at Position (0,0) for $H_2$/He mixture; (c) Substituting Xe for He removed virtually all line broadening of the $H_\alpha$ lines away from the electrodes. Even between the electrodes, broadening was only measurable at high $H_2$/Xe ratios (>4) and even then, significantly less than that observed for He/$H_2$ mixtures.

Figure 3. $H_\alpha$ Broadening Mapped along a Longitudinal Axis 2cm Inward from GEC Edge. (a) This plot shows 'Hot' H$\alpha$ intensity as a function of position and power and is qualitatively consistent with findings along any longitudinal cell axis. That is, the relative intensity is highest between the plates and nearly symmetric for power of 100 W or less, and asymmetric with the highest intensity at the gas input end for power of 150 W or more. (b) Along any longitudinal GEC axis very little change in the magnitude of line broadening as a function of position is found. Also notable: the kinetic temperature of the hot H atoms is higher at an input of 100 W than for an input of 150 W.

Figure 4. Balmer $H_\beta$ Line Broadening. All Balmer series lines studied. (e.g. $H_\alpha$, $H_\beta$, $H_\gamma$, $H_\delta$ at position (3)) were broadened (see Table I).

Figure 5. Fitting of $H_\alpha$ Peaks Using two Gausian Profiles. The sample spectrum shown was taken for a $H_2$/He ratio of 1 x $10^{-5}$ at Position -15,3. This figure is very representative of the fitting process employed in all cases.

Figure 6. Voltage Measured Between Electrodes. This data shows that there is a significant change in voltage between the electrodes as a function of power absorbed by the plasma.

Figure 1a

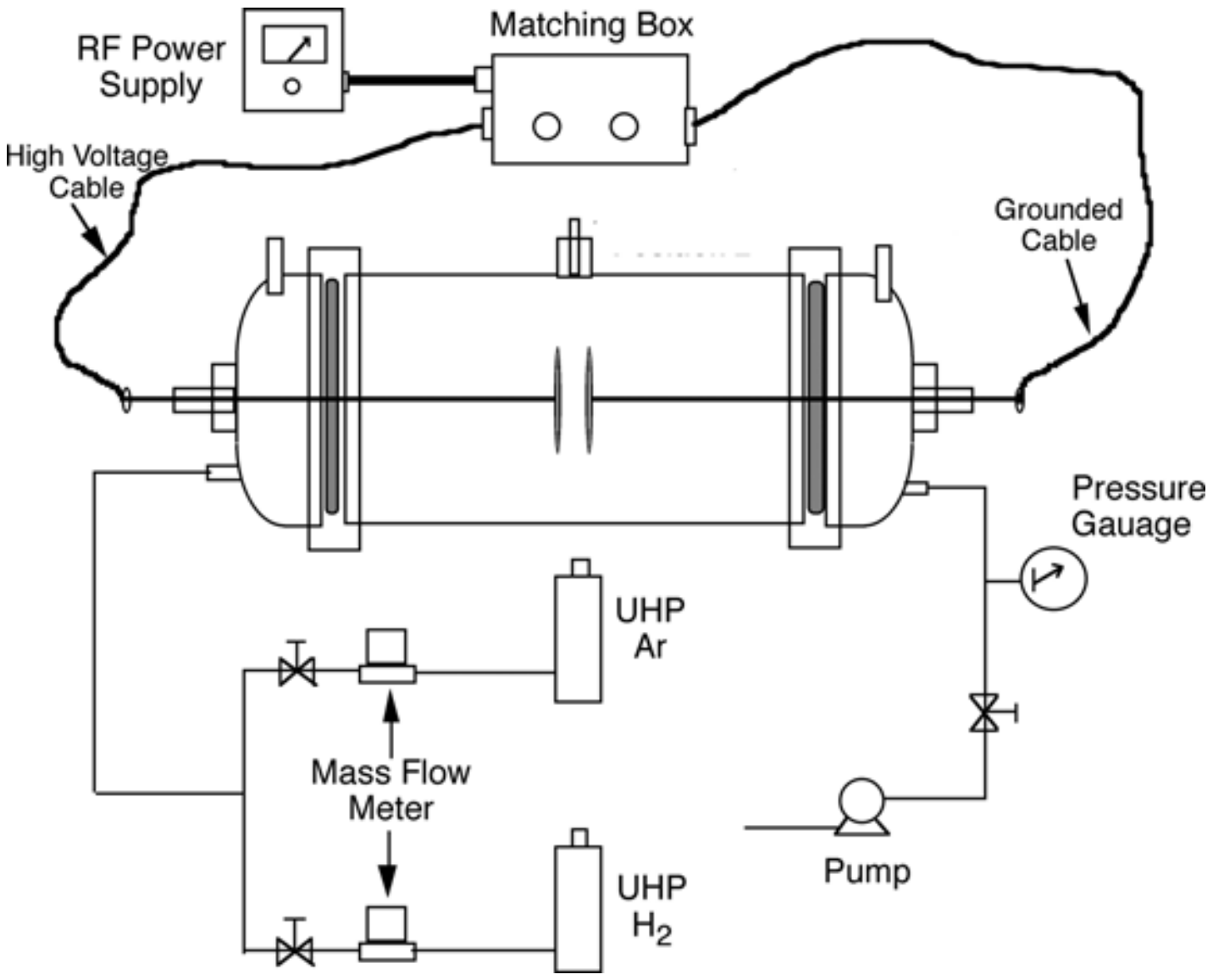

Figure 1b

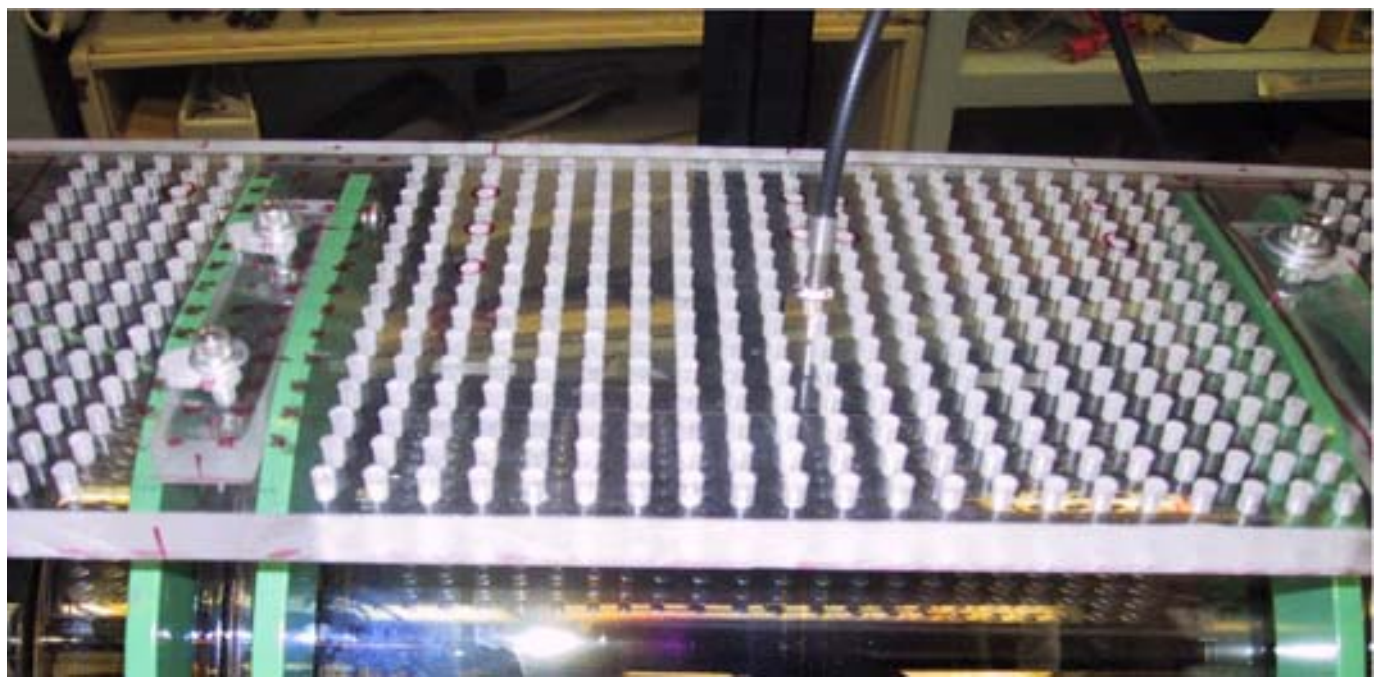

Figure 2a

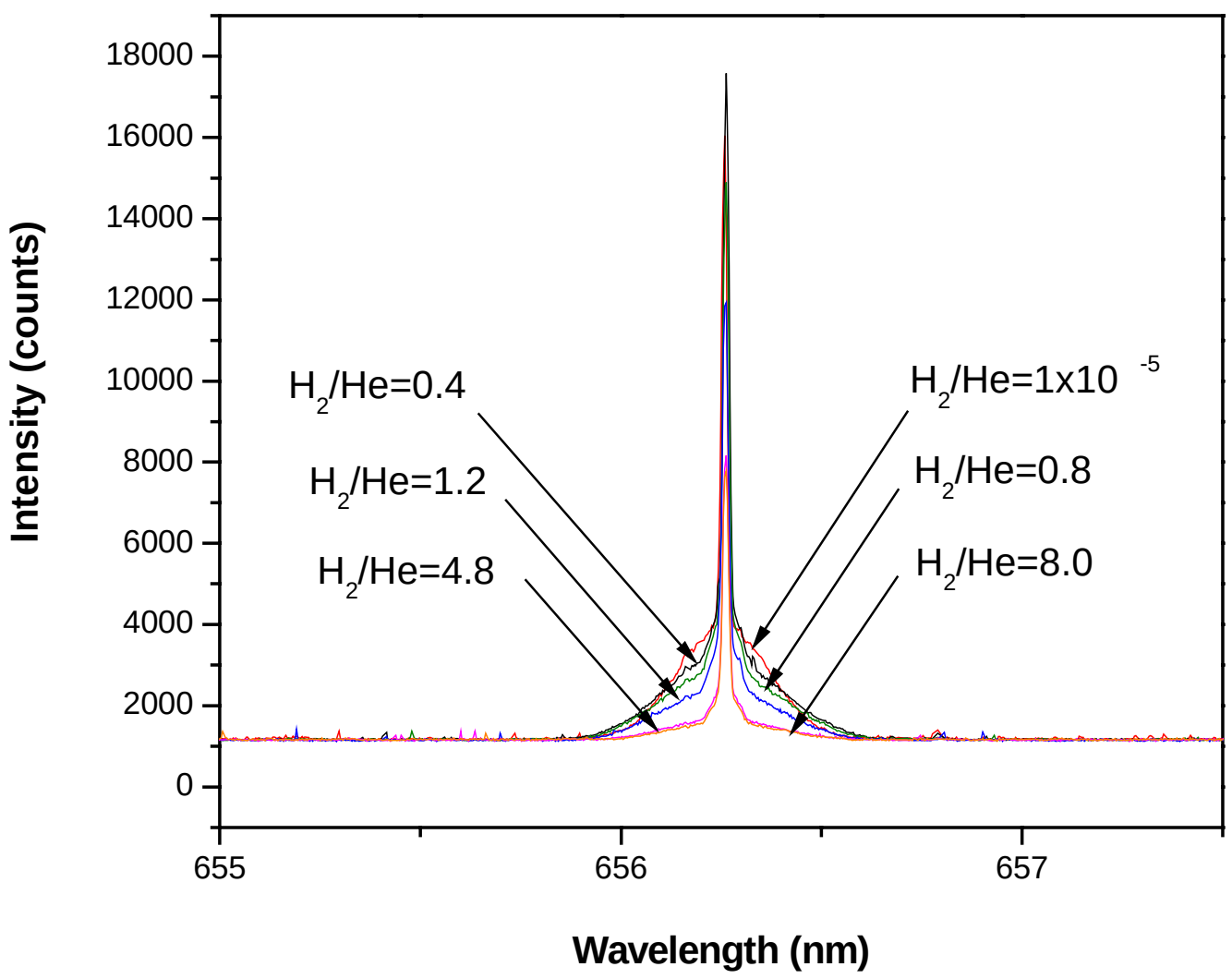

Figure 2b

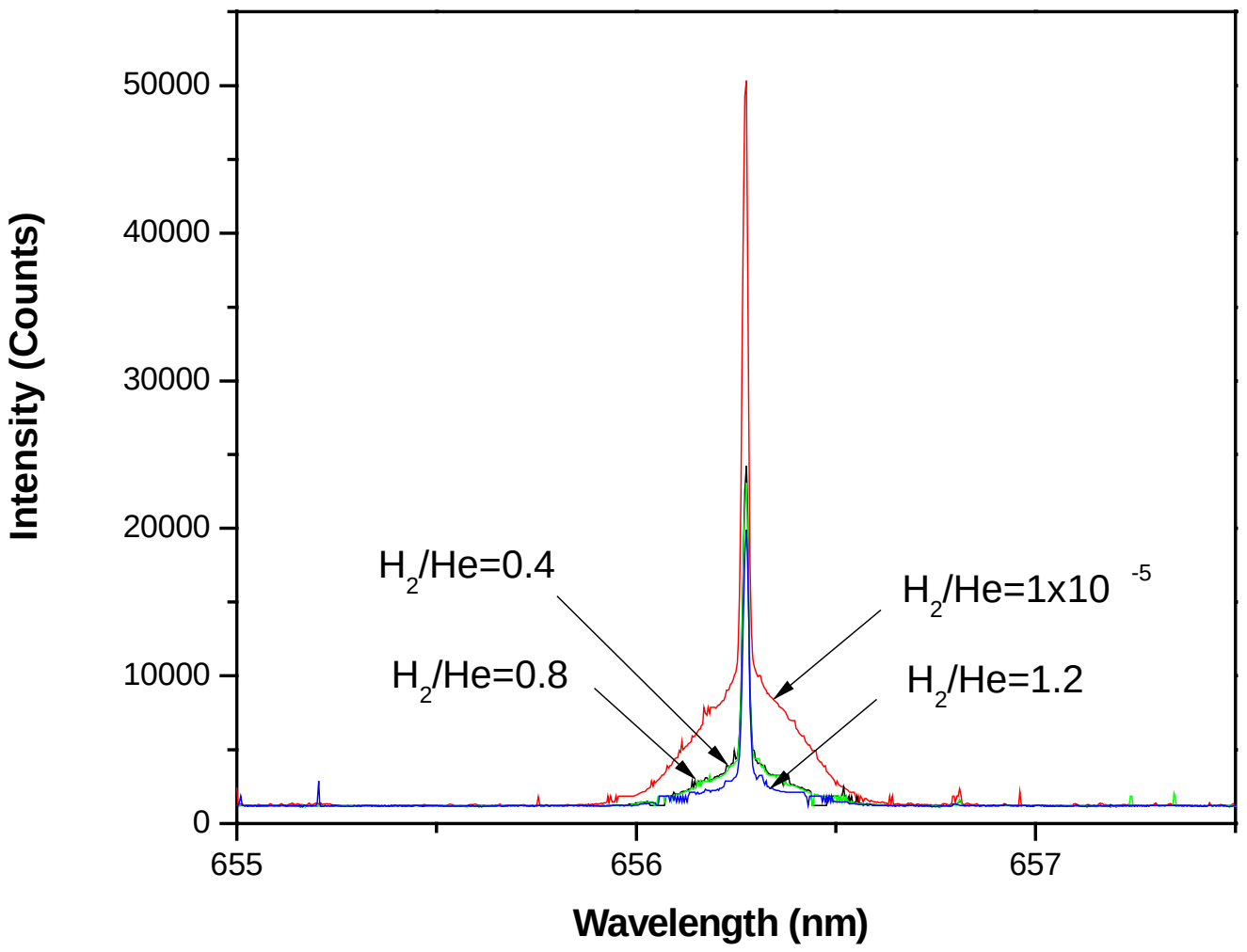

Figure 2c

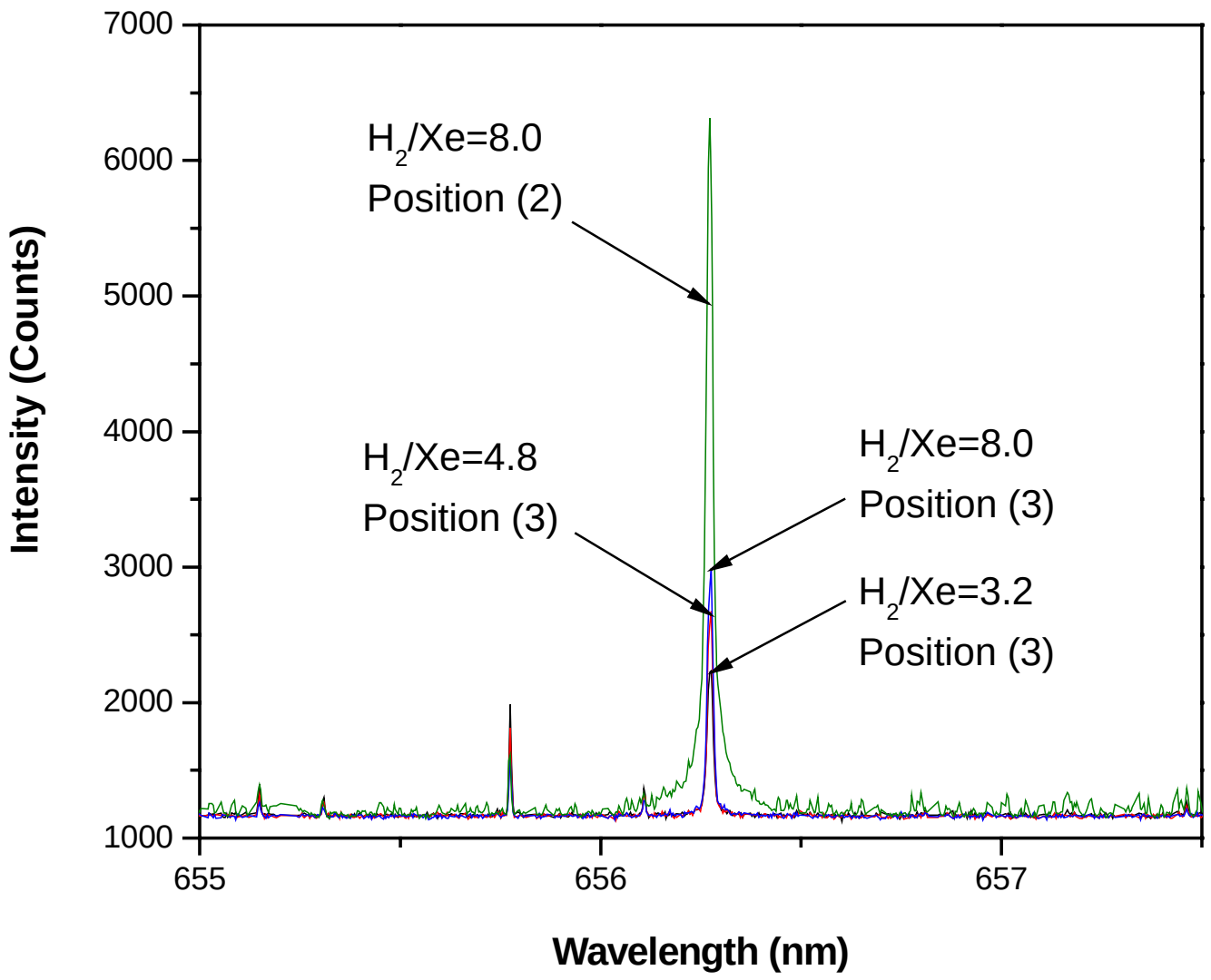

Figure 3a

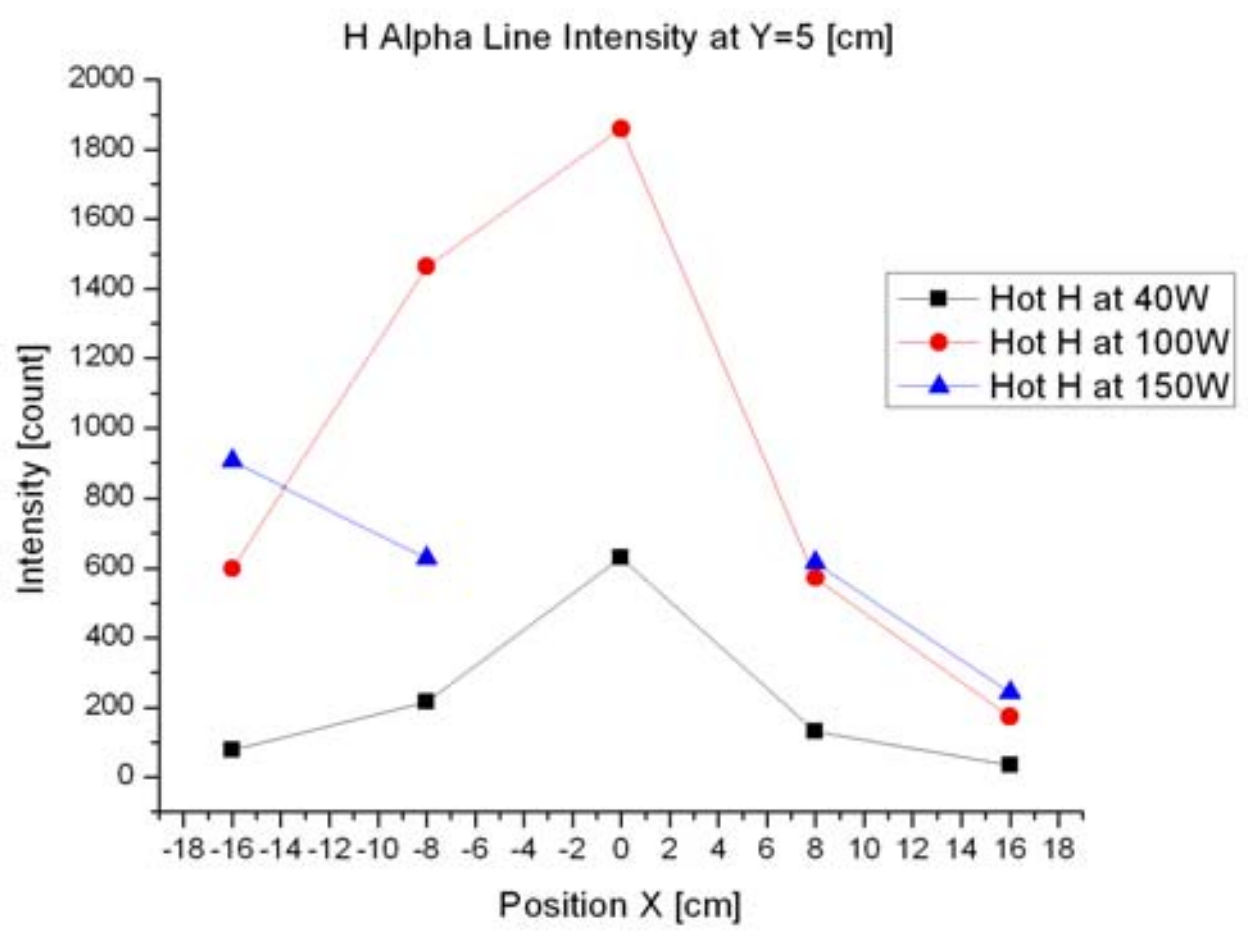

Figure 3b

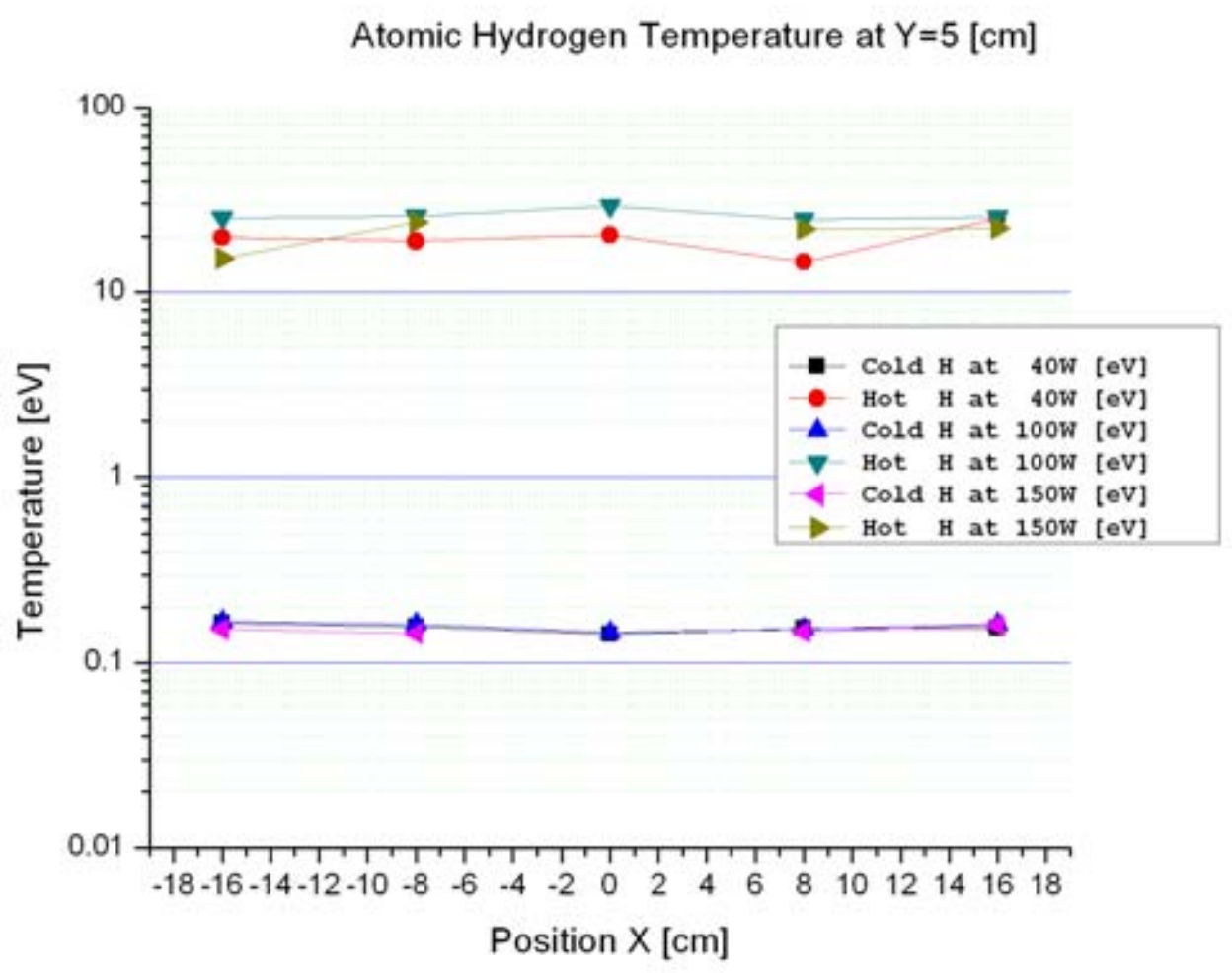

Figure 4

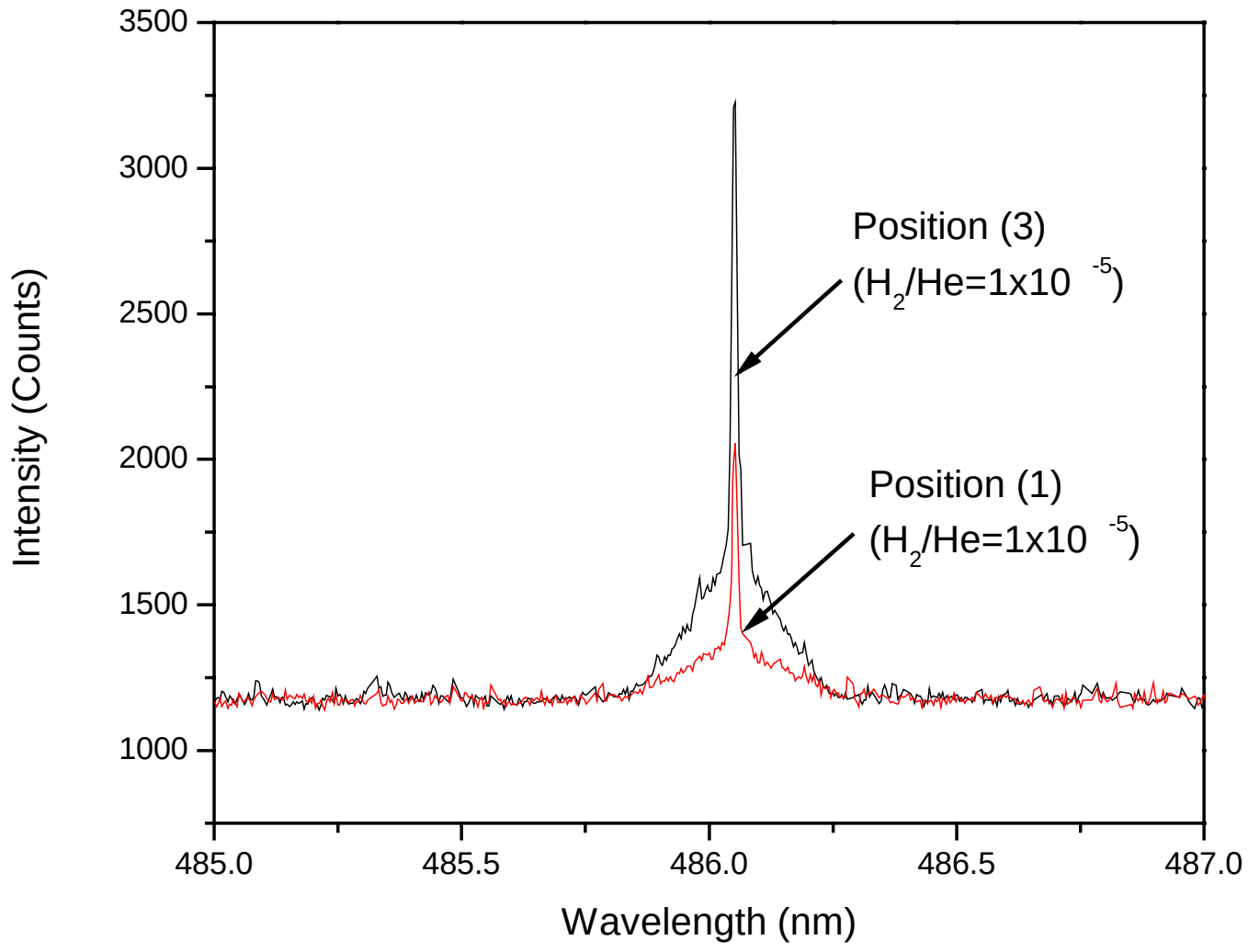

Figure 5

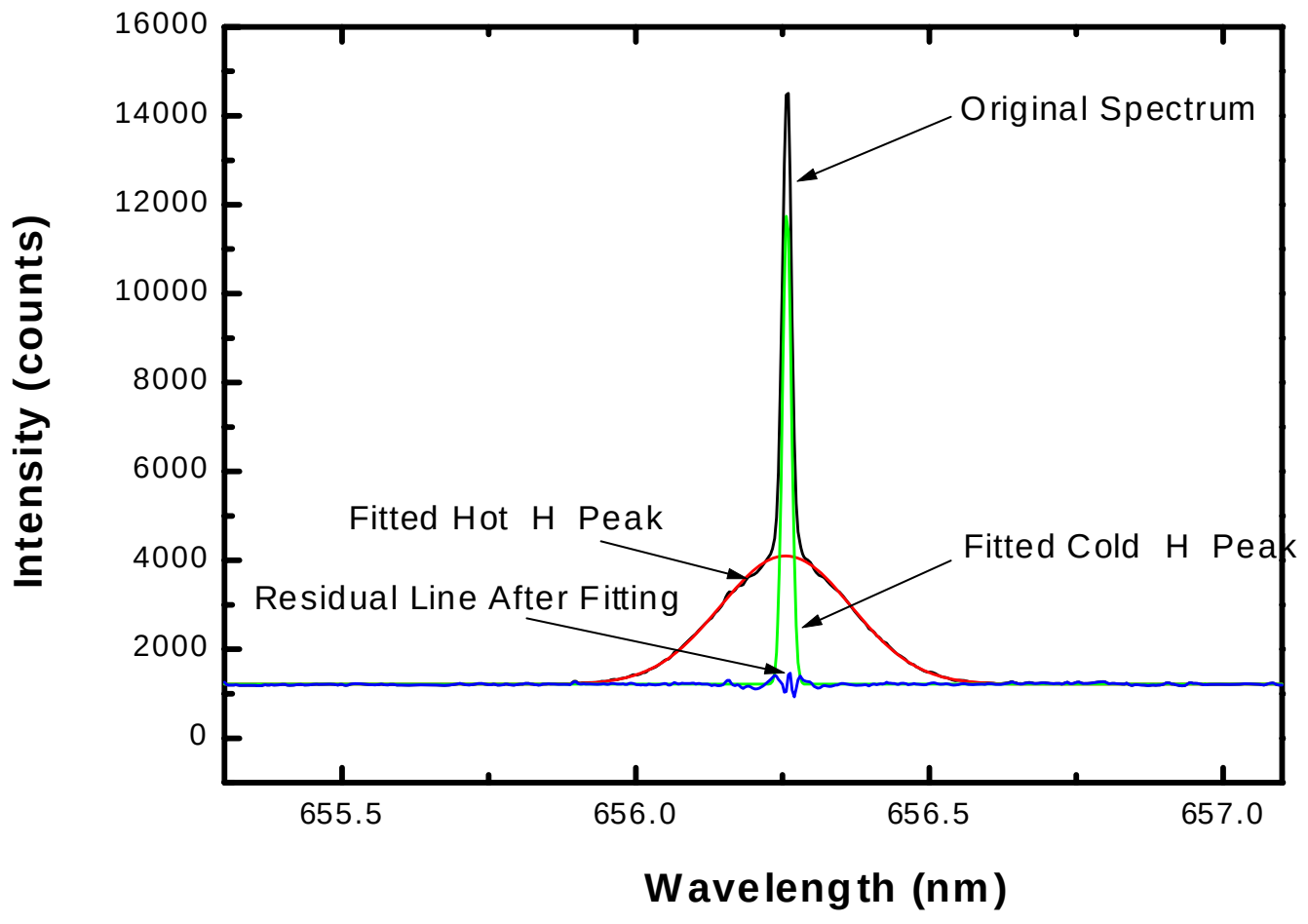

Figure 6

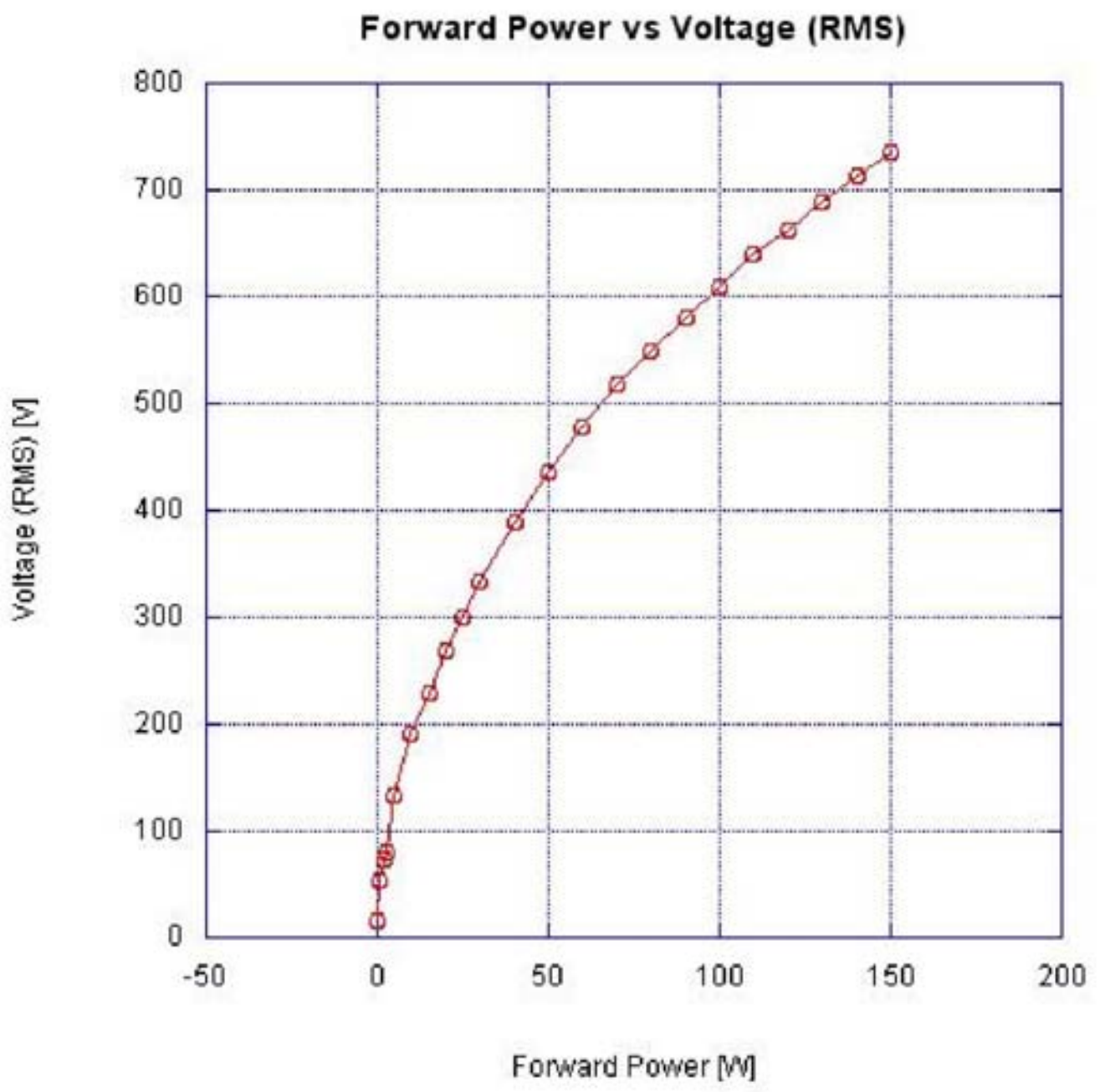